\documentclass[journal,letterpaper,onecolumn,10pt]{IEEEtran}
\usepackage{cite}
\usepackage{color}
\usepackage{epsf}
\usepackage{epsfig}
\usepackage{graphicx}
\usepackage{graphics}
\usepackage{epstopdf}
\usepackage[small]{caption}
\usepackage{amsmath}
\usepackage{amssymb}
\usepackage{amsthm}
\usepackage{amsxtra}
\usepackage[ruled]{algorithm2e}
\usepackage{algorithmic}
\usepackage{enumerate}
\usepackage{multirow}
\usepackage{enumerate}
\usepackage{amssymb}
\usepackage{lipsum}
\usepackage{setspace}
\usepackage{adjustbox}
\usepackage{multirow}
\usepackage{subcaption}
\usepackage{url}
\usepackage{color,soul}
\usepackage{geometry}
\usepackage[utf8]{inputenc}
\usepackage{array}
\usepackage{makecell}
\usepackage[normalem]{ulem}
\usepackage[square,authoryear]{natbib}

\newtheorem{remark}{Remark}

\usepackage{rotating} % adde by Wayes Tushar
\usepackage{graphicx} % added by Wayes Tushar
\newcommand{\Rmnum}[1]{\expandafter\@slowromancap\romannumeral #1@}

\makeatother
\begin{document}
%\doublespace
%
%\title{Evaluation of Recent Technological Developments for Feasible Practice of Negawatt Trading}
\title{Roles of Retailers in the Peer-to-Peer Electricity Market: A Single Retailer Perspective}
\author{Wayes Tushar$^{1,*}$, Chau Yuen$^2$, Tapan Saha$^1$,  Deb Chattopadhyay$^3$, Sohrab Nizami$^1$, Sarmad Hanif$^4$, Jan E Alam$^4$, and H. Vincent Poor$^5$\\$^1${The University of Queensland, St Lucia, QLD 4072, Australia}\\$^2${Singapore University of Technology and Design, 8 Somapah Road, Singapore 487372}\\$^3${The World Bank, Washington, D.C., USA}\\$^4${Pacific Northwest National Laboratory, Richland, WA 99354, USA}\\$^5${Princeton University, Princeton, NJ 08544, USA}
\thanks{$^*$Corresponding author: School of Information Technology and Electrical Engineering, The University of Queensland, St Lucia, QLD 4072, Australia.}
\thanks{\emph{Email address:} w.tushar@uq.edu.au (W. Tushar), yuenchau@sutd.edu.sg (C. Yuen), saha@itee.uq.edu.au (T. Saha), s.nizami@uq.edu.au (S. Nizami), sarmad.hanif@pnnl.gov (S. Hanif), jan.alam@pnnl.gov (J. E. Alam), poor@princeton.edu (H. V. Poor).}}
\IEEEoverridecommandlockouts
\maketitle
\doublespace
% ================================
\begin{abstract}
Despite extensive research in the past five years and several successfully completed and on-going pilot projects, regulators are still reluctant to implement peer-to-peer trading at a large-scale in today's electricity market. The reason could partly be attributed to the perceived disadvantage of current market participants like retailers due to their exclusion from market participation - a fundamental property of decentralised peer-to-peer trading. As a consequence, recently, there has been growing pressure from energy service providers in favour of retailers' participation in peer-to-peer trading. However, the role of retailers in the peer-to-peer market is yet to be established as no existing study has challenged this fundamental circumspection of decentralized trading. In this context, this perspective takes the first step to discuss the feasibility of retailers' involvement in the peer-to-peer market. In doing so, we identify key characteristics of retail-based and peer-to-peer electricity markets and discuss our viewpoint on how to incorporate a single retailer in a peer-to-peer market without compromising the fundamental decision-making characteristics of both markets. Finally, we give an example of a hypothetical business model to demonstrate how a retailer can be a part of a peer-to-peer market with a promise of collective benefits for the participants.
\end{abstract}
 % ====================================
\section{Introduction}\label{sec:background}A retailer is an entity that sells commodities to the public in relatively small quantities for use or consumption. In the electricity market, the concept of an electricity retailer was introduced in the early nineties when some countries started the deregulation process for the power industry by unbundling the vertically integrated power system~\citep{JinZhong_Book_2018}. The purpose was to serve small consumers who could not directly participate in the wholesale market and has been an essential element for ensuring the distribution of electricity to them. However, recently, the involvement of retailers has been de-emphasized in some emerging locally distributed market proposals such as peer-to-peer (P2P) trading~\citep{Tushar_AE_Jan_2021}.

P2P trading is a decentralized electricity management technique that can facilitate the independent decision-making process of prosumers to trade their electricity with one another within a community. Prosumers can be formally defined as the consumers in a community who also have electricity production capability~\citep{Thomas_Nature_2018} by using distributed energy resources such as rooftop solar, battery, and electric vehicles. A fundamental property of decentralized P2P trading is that it does not need a retailer for facilitating the trading of electricity within a local electricity market~\citep{Soto_AE_Jan_2021}. This lack of influence from retailers has been shown to be very effective in enabling extensive participation of prosumers in P2P trading to reduce greenhouse gas emission within the electricity network, provide demand flexibility and electricity services to the grid, and to ensure greater prosumer privacy. This lack of influence from retailers has been shown to be very effective in enabling extensive participation of prosumers in P2P trading  to reduce greenhouse gas emission within the electricity network, provide demand flexibility and electricity services to the grid, and ensure greater prosumer privacy. For example, see the recent survey articles on decentralized P2P energy trading research and pilot demonstration in \citep{Tushar_AE_Jan_2021,Thomas_Nature_2018,Sousa_RSER_Apr_2019}, and \citep{Tushar_TSG_2020_Overview}.

However, excluding retailers from the market also hinders the potential large-scale implementation of P2P trading in today's electricity market for a number of factors. Firstly, operating under a new regulatory and retail scheme would require reforming current market structures and electricity policies. This may require huge investment, substantial research and time, and the commitment of existing market stakeholders, which is a challenging task~\citep{Esther_AE_Jan_2018}. Secondly, prosumers with solar may also need electricity from the grid when the available supply of electricity is lower than their demand. Therefore, as customers and prosumers will still need to rely on existing retailers, the participation of retailers in the P2P market is important. Finally, diminishing retailers' interest in P2P markets may increase the price of electricity significantly for the rest of their customers~\citep{Barbose_Report_2017}. Such impacts on the overall electricity sector can potentially disrupt overall market structures and affect the sustainable operation of P2P markets into the future. An example of previous electricity management schemes being discontinued under similar circumstances can be found in \citep{FiTExplained_2016}.

Given this context, some electricity service providers are questioning this fundamental characteristic of P2P electricity trading and have started to defend the involvement of retailers within the P2P framework as an essential element of the network~\citep{enosi_whitepaper2019}. However, not many studies have challenged this fundamental circumspection of P2P trading. Therefore, a clear insight of how the incorporation of a retailer within the framework may impact the overall electricity trading behavior of the prosumers is yet to be established. In other words, it is important to discover whether a retailer can be incorporated into a P2P market without compromising the independent decision making properties of the participating prosumers and help them to improve their overall revenues. Or, will such an incorporation could reduce prosumers' benefits compared to standard P2P trading schemes?

This paper takes the first step to discuss a framework that is capable of integrating a retailer within a P2P market without compromising the basic properties of both markets. In doing so, we briefly provide a short background on the properties of both P2P and retail-based electricity markets followed by a proposition on the expected properties of a retailer-incorporated P2P market. Then, we propose a framework of a hypothetical P2P market that can be utilized to include a retailer within the P2P market and discuss how the proposed framework can be utilized followed by some concluding remarks.

It is important to note that the proposed market model is different from the models using community-based peer-to-peer sharing or aggregator-based trading. In community peer-to-peer energy sharing models, such as in the Brooklyn Microgrid~\citep{Esther_AE_Jan_2018}, prosumers communicate with the community manager for sharing energy with one another. The community manager acts as a coordinator without directly influencing the decision making process of the prosumers where prosumers share limited information with the community manager for maintaining a higher degree of privacy and security~\citep{Tushar_AE_Jan_2021,Tushar_TSG_2020_Overview}. However, the community manager does not enable prosumers to participate in the spot market for trading their energy after meeting the local needs via peer-to-peer trading. In an aggregator-based trading model, on the other hand,  the aggregator has a contract with each participating household specifying the amount of energy that each household needs to contribute to the VPP and the aggregator is in control of different households to procure their agreed-upon amounts \citep{Siano_Aggregator_SJ_2019}. However, this model restricts the decision-making flexibility of prosumers in terms of when they want to participate in the spot market and how much they would like to contribute for that purpose. The retailer-incorporated peer-to-peer sharing model proposed in this perspective overcomes these limitations by enabling prosumers with energy surplus to participate in the spot market and make extra revenue without the need to compromise their flexibility in choosing the energy amount they would like to trade in the spot market and the schedule of trading. Further, it also enables multiple retailers to participate and reap benefits by providing services to the prosumers. A summary of the key differences between the proposed retailer-based P2P and community-based P2P and aggregator-based trading models is shown in Table~\ref{table:Diff}.
\begin{table*}[t]
\centering
\caption{Difference between the proposed retailer-based P2P and community-based P2P and aggregator-based trading models.}
\small
\begin{tabular}{|m{2cm}|m{2cm}|m{4cm}|m{4cm}|m{4cm}|}
\hline
\textbf{Market model} & \textbf{Example} & \textbf{Prosumers' participation in the spot market} & \textbf{Flexibility of prosumers decision-making} & \textbf{Participation of retailers in the trading}\\
\hline
Community-based P2P & \cite{Esther_AE_Jan_2018} & No & Flexible & No\\
\hline
Aggregator-based trading & \cite{Siano_Aggregator_SJ_2019} & Yes & Restricted & No\\
\hline
Retailer-based P2P & \emph{Proposed model} & Yes & Flexible & Yes\\
\hline
\end{tabular}
\label{table:Diff}
\end{table*}

This paper is most suitable for readers who have a basic understanding of energy trading, energy markets, and research on future power systems. Meanwhile, readers who do not have this necessary background can also easily understand the content after some preliminary reading about P2P energy trading \citep{Tushar_TSG_2020_Overview}, Federated Power Plants (FPP)~\citep{Thomas_Nature_2018}, and energy markets \citep{Kirschen_Book_2019}.
\section{Brief Background}\label{sec:background}
\begin{figure}[t!]
\centering
\includegraphics[width=0.7\columnwidth]{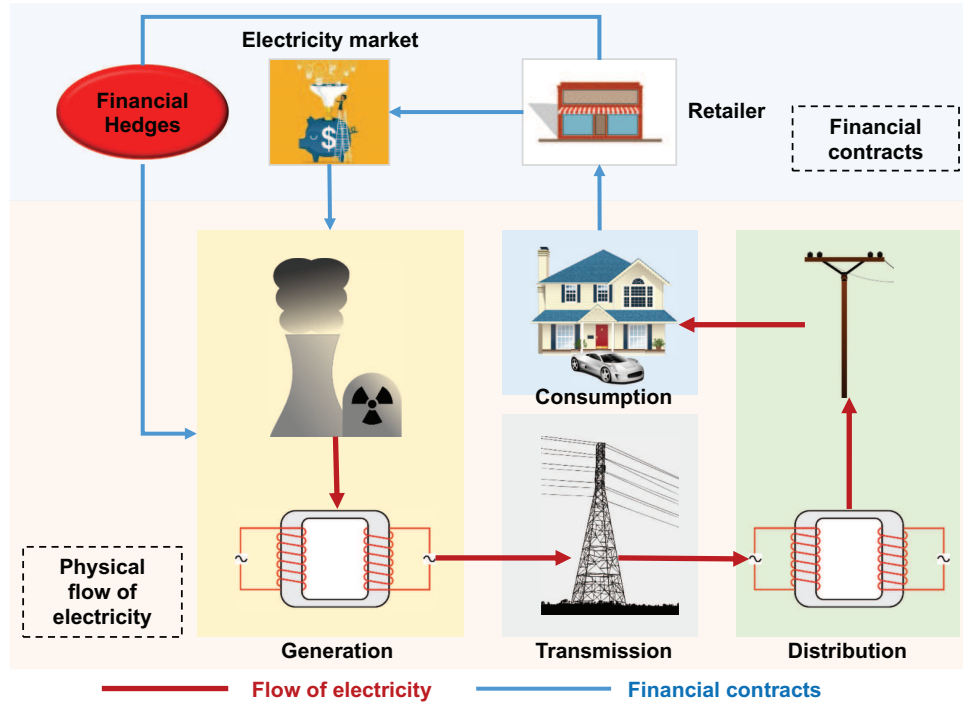}
\caption{This figure shows an overview of retailer-based electricity market. According to this figure, the electricity is generated in a power station and transmitted over the transmission line to the substation. From the substation, electricity is distributed among customers for consumption. Here, the retail-based electricity market is responsible for the trading of electricity for the customers where the financial contracts to facilitate this flow of electricity are done between generators and retailers as well as between the retailer and the customers through the electricity market. All the images used in this figure have been taken from royalty free image website: {https://pixabay.com/}.}
\label{Fig:RetailMarket}
\end{figure}
\subsection{Retail-based market}Retail electricity markets are the final step in the electricity supply chain and an electricity retailer can be formally defined as a stakeholder in the national electricity market that buys electricity from the wholesale markets and sells it to customers~\citep{Kirschen_Book_2019}. A retailer acts as the main interface between the electricity generators and customers. It offers electricity services to customers, which allows it to sell electricity, gas, and electricity related services to residents and business owners. A retail market also promotes retail competition and allows customers to choose between competing retailers to receive appropriate electricity services. Such competitive retail markets with appropriate customer protections open opportunities for innovation, product choice, and competitive pricing. As a part of the retail market, each retailer has to perform some other common functions in order to compete effectively. Examples of such functions include billing, credit control, customer management, distribution of use-of-system contracts, reconciliation agreements, spot market purchase agreements, and hedge contracts~\citep{AEMC_2021}.

In traditional retail markets, a retailer charges its customers for providing electricity. The amount of the bill depends on electricity usage, access fee, and service charge. The access fee covers the cost to the retailer for using the existing transmission and distribution network to supply electricity to the customers. The service charge, on the other hand, includes the marginal cost of transferring electricity through the network. Meanwhile, with the advancement of event-driven service-oriented architectures~\citep{LinaLan_IJOBE_2015}, electricity retail markets are facing an unprecedented change. Now, electricity users can bid for power based on the demand of each device within their households and how much they want to pay. Generators or electricity suppliers can also bid automatically based on the costs for supplying electricity. Suppliers can forecast return-on-investment through real-time energy market analysis, which also opens opportunities for retailers to become parts of P2P electricity markets by facilitating the trading of electricity by prosumers. A demonstration of retail-based electricity market is shown in Figure~\ref{Fig:RetailMarket}.
\begin{figure}[t!]
\centering
\includegraphics[width=0.5\columnwidth]{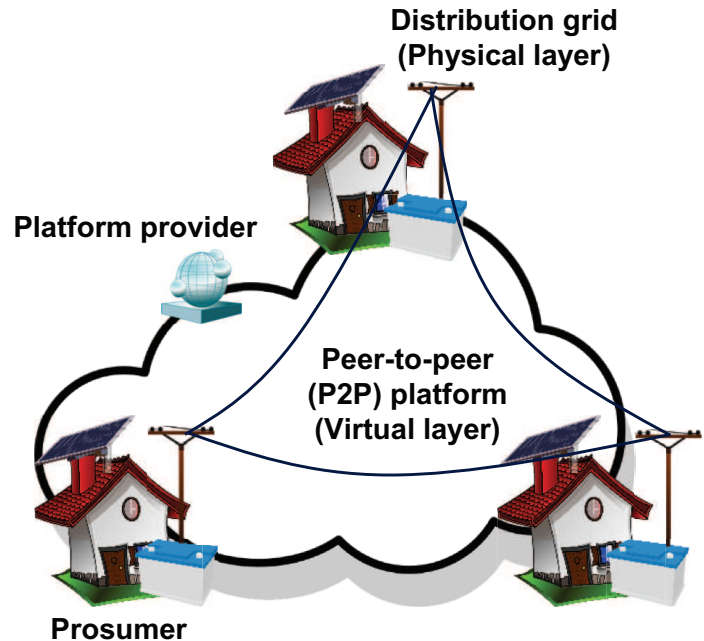}
\caption{This figure shows a model for a P2P electricity trading market, in which each peer of the network is connected to one another through a cloud P2P platform provided by a third party. While all the financial settlements and negotiations are conducted in the virtual layer (P2P platform), the actual transfer of electricity between prosumers is done through the physical layer of the network, that is, the distribution network. All the images used in this figure have been taken from royalty free image website: {https://pixabay.com/}.}
\label{Fig:P2PMarket}
\end{figure}

\subsection{P2P market}In a P2P market, as shown in Figure~\ref{Fig:P2PMarket}, two peers can agree to trade a certain amount of electricity (or, the right of buying electricity, that is, negawatt~\citep{Nature_energy_Wayes_2020}) with one another for a price negotiated and set by them without any centralized supervision. Participants can rely on bilateral economic dispatch in which consumers of the electricity can express their preferences on consuming the local electricity \citep{Sorin_TPWRS_Mar_2019}. Prosumers may also consider upstream-downstream electricity balance and market uncertainly to set their preferences for trading electricity within a P2P market \citep{Thomas_TSG_Mar_2019}.

The heart of a P2P market is the optimization that explicitly defines individual problems of each prosumer while preserving their privacy. The examples of two types of P2P trading platform follows. In a fully decentralized P2P trading, each prosumer uses suitable optimization to share the power, negawatt, and price that it is willing to trade, without revealing any private information. In a community-based P2P market, on the other hand, each prosumer trades anonymously with one another within the community, as it is centrally handled by the community manager. Hence, the negotiation process can be solved in a distributed manner, in which each prosumer solves its own problem but shares some required information with the community manager. Indeed, if some of the prosumers have surplus electricity after sharing within the local P2P market, they may choose to form an FPP~\citep{Thomas_Nature_2018} - a concept similar to a virtual power plant - for trading in other alternative markets, for example, in a retail or spot market. A more detailed description of P2P markets can be found in \citep{Sousa_RSER_Apr_2019}. 
\subsection{Properties of a retailer-based P2P market}Given the characteristics of P2P and retail-based electricity markets, a retailer-based P2P market should be able to serve two types of customers: 1) prosumers within who need to buy their required electricity from the retailer-based market at times when their need cannot be fulfilled by either their own generation or electricity from the local P2P market~\citep{Ableitner_AE_Jul_2020}, and  2) prosumers who actively participate in P2P trading for exchanging electricity among themselves to balance their demand and supply and are interested to form a FPP to trade their surplus electricity in the spot or retail market. Therefore, it is necessary for a retailer-based P2P market to possess some selected properties of both retail-based and P2P markets as discussed earlier. In particular, in a retailer-based P2P market, the independence of each individual decision-making process should not be affected by the presence of the retailer and prosumers should be able to express their preferences on electricity management. Further, prosumers should have the ability to use suitable optimization techniques to share the electricity and price information with one another to negotiate for trading. The negotiation process between prosumers should be in a distributed manner with limited information sharing with the retailer, if necessary. 

Meanwhile, the retailer should be able to act as a facilitator of electricity trading between the suppliers and consumers within the network. It should also be able to provide other services such as billing, customer engagement and participation in the spot/retail market as a FPP, reconciliation agreement, and hedge contracts within the market and reap monetary revenue from the prosumers for providing such services. Of course, an important question that need to be considered is that whether or not a retailer would be motivated to be a part of P2P markets. To that end, we note that\footnote{See the white paper released by enosi in 2018:\url{https://drive.google.com/file/d/1OWTxITLOCx6mxchr6pR0sX538uR-wDsk/view?usp=sharing}.} the average cost for an electricity retailer to serve its customer is \$80 per annum;  in general, electricity retailers have very poor customer satisfaction ratings; and the installation of typical 6 kW rooftop solar in consumers premises has reduced the revenue for electricity retailers by about 43\%. Therefore, it would be very unlikely for a retailer to dismantle its customers base and not to perform additional roles and participate in P2P markets. As such, to articulate how a P2P trading mechanism can successfully capture these properties, we discuss a market structure in the next section.
%
% -------- Figure ---------%
\begin{figure*}[t]
\centering
\includegraphics[width=0.8\textwidth]{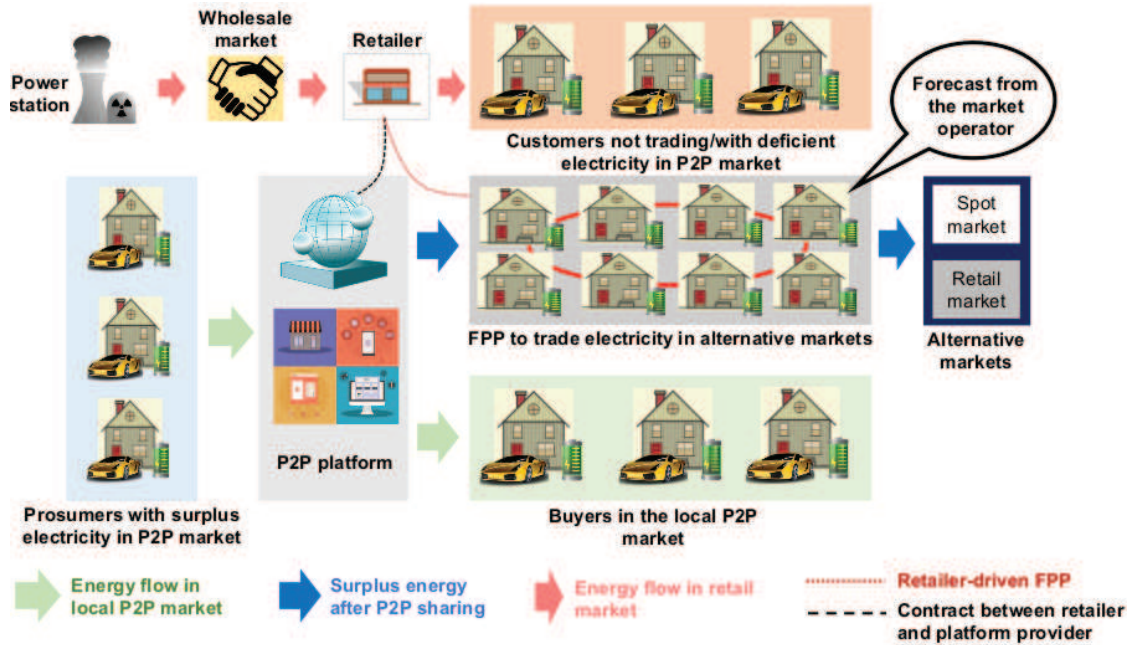}
\caption{This figure illustrates the proposed framework for incorporating a retailer into the P2P market. In this framework, a third party is providing the P2P platform for the prosumers to participate in local electricity trading among themselves whereby the retailer can help prosumers to buy electricity from the retail market, if needed. All the images used in this figure have been taken from royalty free image website: {https://pixabay.com/}.} 
\label{Fig:SingleRetailer}
\end{figure*}
\section{P2P Market with Retailer}
In this section, we discuss a P2P market in which a single retailer is serving its customers within an electricity network. Depending on who is providing the P2P platform for the electricity trading service, we propose that there could be two possible scenarios. In the first scenario, as shown in Figure~\ref{Fig:SingleRetailer}, a third party is responsible for providing the P2P platform to the prosumers for locally trading electricity among themselves. There is a retailer, who can help the prosumers with energy deficiency to buy electricity from the retail market, if needed. There is a contract between the retailer and the third party that mandates that each prosumer needs to choose the retailer designated by the third party for its electricity need in order to use the P2P platform provided by the third party. Prosumers can form a FPP by using the P2P platform and trade their surplus energy, if appropriate, to other alternative markets through the retailer. Thus, prosumers rely on the retailer for both participating in local P2P market as well as trading their surplus energy as a FPP to the spot/retail market. In the second scenario, it is considered the retailer also owns the P2P platform. All prosumers that are interested to participate in P2P trading need to sign-up with the retailer for seeking this additional service with a supplementary subscription fee per month, whereas the other aspects remain same as the first scenario. 

Thus, involving a retailer within the P2P market will enable prosumers with enough electricity surplus to manage and trade their electricity in two different markets including the local P2P market and either the spot or retail market at any given time. Subsequently, prosumers not only will be able to obtain more monetary benefit through the proposed scenario, but also make noticeable contribution to environmental sustainability through increasing the flow of clean energy within the electricity network. 

\subsection{Electricity trading in the local market}
Like existing P2P market models~\citep{Sawwas_AE_Feb_2021}, in the local electricity market, prosumers can share their electricity with one another. In doing so, each prosumer will first meet its own demand using its renewable electricity resources such as rooftop solar and battery. When the electricity from its own resources is insufficient, it will then meet the demand by buying electricity from the other participants' available solar surplus. If the net surplus of the other participants' is insufficient, then participants will meet their demand by buying P2P electricity from the other participants' battery charge. In cases where the available electricity from other participants is insufficient, a buyer participant will need to buy its remaining electricity from the retailer with a standard market mechanism~\citep{JinZhong_Book_2018}. Thus, in the local market, the P2P trading platform will identify each prosumer either as a seller or a buyer of electricity depending on their available generation and demand after meeting their needs from own renewable electricity resources. Each seller bids its price per unit of electricity and the amount of electricity that it would like to sell from its preferred range into the market. Here, preferred range refers to the maximum and minimum values of electricity and price from which a seller can choose to bid in the market for trading purposes. Similarly, each buyer submits its preferred price and the electricity that it wants to buy through the platform.

As an electricity service provider of the P2P trading market, the retailer will facilitate the electricity trading by determining whether the total electricity bid by all sellers are enough to meet the demands of all buyers at the market clearing price, and advising the participants to re-bid if necessary. Further, the retailer will also establish an appropriate market clearing price, for example, double auction~\citep{Kaixuan_AE_May_2019}, mid-market rule~\citep{Raisul_AE_Mar_2019}, modified auction~\citep{Zhang_Energies_Sep_2020}, or game theory~\citep{Paudel_TIE_2019}, to facilitate the flow of electricity from sellers to buyers. If needed, the retailer will also fulfil the electricity demand of the selected buyers, if needed, through the traditional retailer-based market mechanism.

\subsection{Electricity trading in the spot market}
While the P2P trading of electricity in a local market discussed in the previous section is similar to those studied in the existing literature, the proposed trading model in this perspective contrasts existing studies by further enabling prosumers to trade their electricity in the spot market through the involvement of the retailer. That is, once prosumers finish trading their electricity at the local market, the retailer can expedite the bidding of remaining surplus electricity in prosumers' batteries, as a FPP, in the spot or retail market. For a medium size community, the overall electricity surplus being injected into the grid is low. Therefore, it can be assumed that this will not affect the supply and demand balance, and will have negligible impact on the market clearing price. Thus, the FPP is purely a price taker, and for the bidding, the retailer will only consider the quantity of bids, not the prices. As shown in Figure~\ref{Fig:SingleRetailer}, the decision to bid in either the spot or retail market is determined by considering the spot market forecasts for the current dispatch interval as given by the electricity market operator in a country~\citep{AEMO_MarketForecast}. If the forecast for the spot market exceeds the presumably constant retail selling price, then bids will be allocated to the market with the highest forecast. Otherwise, bids will be allocated to the retail market.

When the retail market is preferred, on the other hand, it is assumed that the retailer sells the electricity from the participants to other customers at a constant retail price. This price is used as the determining factor for bidding in the spot market and the retail market. Since the retail price is constant, the determination of a bidding quantity each dispatch interval does not rely on the forecast if retail selling is preferred. On the other hand, if the spot market forecast is greater than the retail market price, then a portion of the excess surplus of the FPP will be bid in the market with the highest forecast. Nonetheless, surplus electricity from the FPP can only be bid into either the retail market or the spot market in a dispatch interval, depending on the value of the forecast of the spot market compared to the constant retail price. Once the market has been chosen, the quantity to bid is determined based on a pre-defined spot and retail bidding quantity equations. The revenue created by the FPP for bidding is calculated using a constant retail price and the spot prices given by the electricity Market Operator of the region, for example, Australian Energy Market Operator in Australia~\citep{AEMO_MarketForecast}. The revenue that a participating prosumer generates for the FPP can be calculated based on the participant's contribution to the overall bidding quantity for that interval.

\section{Further discussion} In the proposed model, the total profit earned by each prosumer by participating in the P2P trading stems from the benefit of trading in the local market and the additional revenue that is received from the retailer to bid its surplus electricity in the spot market as a part of the FPP. On the other hand, the retailer earns its revenue from both the subscription fee of the prosumers that they pay for taking the additional service of combined bidding and the profit of bidding FPP's energy into the spot or retail market. Indeed, the retailer needs to offer a sufficient percentage of its additional benefit to the participating prosumers so that they are active in the FPP for bidding surplus energy. To fairly decide on the percentage of profit that the retailer needs to share with each prosumer can be determined using any standard fair resource allocation distribution scheme~\citep{Saad-coalition_2009}. 

\subsection{Demonstration through a toy example}A simple toy example can be used to demonstrate this. Let us assume a community consisting of 10 prosumers who participate in P2P energy sharing with one another. Each prosumer has a 6kW solar system installed within its premises and based on its demand and solar generation, it can either act as a seller or a buyer at any given time slot. Each prosumer has access to the spot market price forecast of electricity market authority, which is reasonably accurate with occasional missing very few price spikes. Now, in this example, we show how the proposed model will work for both prosumers and the retailer at a given time slot when (1) the spot market price is larger than the retail price with accurate forecast, (2) the spot market price is larger than the retail price with an inaccurate forecast, (3) when spot market price is higher but inaccurate forecast shows its lower than the retail market, and (4) the spot market price is lower than the retail market price. For this purpose, we assume that there are 5 prosumers with a total of 15 kWh surplus that they can either sell to the spot market or the retail market. For simplicity, it is further assumed that each prosumer has a 3 kWh contribution to that total 15 kWh surplus. Here is it important to note that, in general, generation capacities of household customers are very low compared to the lowest threshold of energy volume that is needed by a prosumer to participate in the spot market~\citep{Faia_Energies_Jun_2021}. Hence, a prosumer cannot participate in the spot market alone. In the proposed framework, a retailer provides a mean for prosumers to trade their surplus energy in the spot market. In Table \ref{Table:ToyExample}, we show how adopting the proposed  retailer-based P2P market can benefit the prosumers and the retailer compared to traditional P2P market by enabling them to opportunistically participate in the spot market.
\begin{table*}
\centering
\caption{This table shows the excess revenue that the retailer and each prosumer can make from the P2P market by participating in the spot market. The retail electricity selling price is assumed to be equal to 7 cents per kWh \citep{Zahedi_RESER_2010} and the spot price for case 1 and case 2 for the considered time slot is assumed to be 800 cents per kWh (the price is chosen from the spot price list of February 10, 2017, in Australian National Electricity Market~\citep{AEMO_NEM_2020}). In case 2, we assume that there is a forecast error by the prosumers. However, since the forecasted price is more than the retail price, the prosumers still sell their excess to the spot market and gain revenue based on the actual market price. For deciding the retailer's benefit for facilitating this trading, in this example, it is assumed that the retailer kept 50\% of the total revenue from the trading and then distribute the rest to the prosumers according to their contribution to the total traded energy. In case 3, although the spot market price is high, due to inaccurate forecast which is lower than the retail price, prosumers participate in the retail market. While this detrimentally reduces their revenue compared to what they could achieve from the spot market, the excess revenue beyond P2P sharing to the prosumer is as same as the traditional P2P market. Case 4 demonstrates the same revenue as the traditional P2P market where prosumers sell their excess energy in the retail market after sharing in the P2P.}
\footnotesize
\begin{tabular}{|m{0.55cm}|m{1cm}|m{1.5cm}|m{1.5cm}|m{1.5cm}|m{0.95cm}|m{0.95cm}|m{1cm}|m{1cm}|m{1.35cm}|m{1.35cm}|m{1.5cm}|}
\hline
\textbf{Case} & \textbf{Total surplus} (kWh) & \textbf{Retail price} (cents/kWh) &\textbf{Spot price} (cents/kWh)& \textbf{Forecast to the prosumer} (cents/kWh) & \textbf{Trade in Retail market} & \textbf{Trade in Spot market} & \textbf{Total revenue from trading} (\$) & \textbf{Revenue to the Retailer} (\$) & \textbf{Revenue to prosumer for proposed mechanism} (\$) & \textbf{Revenue to prosumer for traditional P2P market} (\$) & \textbf{Improvement of revenue per prosumer}\\
\hline
1 & 15 & 7 & 800 & 800 &&$\surd$&120&60&12&0.21&57 times\\
\hline
2 & 15 & 7 & 800 & 400 & & $\surd$ &120 & 60 & 12 & 0.21 & 57 times\\
\hline
3 & 15 & 7 & 800 & 6 & $\surd$ & & 1.05 & 0 & 0.21 & 0.21 & Same performance\\
\hline
4 & 15 & 7 & 0 & 0 & $\surd$ & & 1.05 & 0 & 0.21 & 0.21 & Same performance\\
\hline
\end{tabular}
\label{Table:ToyExample}
\end{table*}

As can be seen from Table~\ref{Table:ToyExample}, the benefit from participating in the proposed scheme is substantial, which is $57$ times more than the traditional P2P scheme in this particular case. The higher spot market price was the main reason for such huge benefit per prosumer. It is important to note that such a higher spot market price is not a random event, rather it can happen on regular basis in a renewable dominated region~\citep{AEMO_NEM_2020}. However, the forecast of the prosumers about the spot market price dictates their benefits from participating in the market using the proposed scheme. The recent advancement in forecasting technique can accurately predict the future events~\citep{Zhang_AppliedEnergy_Jan_2020}. Hence, it can be reasonably assumed that that the prosumers will be able to estimate the change in the future spot market price with a legitimate accuracy and be benefitted by the proposed scheme. Meanwhile, in times when the spot market price is lower than the retail price or the forecast about the spot price is inaccurate, the proposed scheme demonstrates similar benefits to prosumers and the retail as existing P2P schemes. Furthermore, the proposed framework also provides retailers a great way to hedge against market power risk. For example, if the generators exercise market power to shoot wholesale prices up in thousands of dollars per MWh (as in case 1 of the given example) - the proposed retailer-prosumer trade would shield the retailer to some extent and the threat of it would prevent the generators from exercising market power.

\begin{figure}[t]
\centering
\includegraphics[width=0.5\columnwidth]{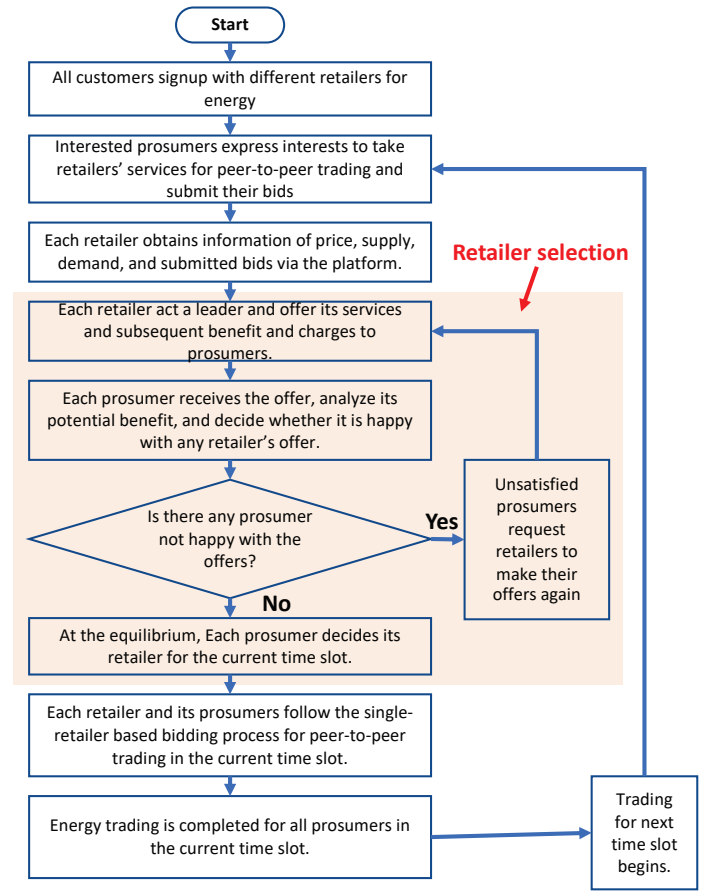}
\caption{This figure demonstrates one example of how multiple retailers can participate in a P2P market and help both retailers and prosumers to improve their revenues from P2P energy trading.}
\label{Fig:MultiRetailer}
\end{figure}

\subsection{Extension to multiple retailers case}The P2P energy trading mechanism can also be extended to the case when there are multiple retailers in the market. However, involving multiple retailers in P2P is challenging considering the modeling of various scenarios that may arise due to their heterogeneous interest and interaction patterns within the market. For example, different retailers may use different pricing mechanisms to charge the customers and have different portfolios of services that they can offer. Retailers may also have different types of energy contracts with renewable and non-renewable energy generators that may also affect their business dealing with different customers with and without renewable energy resources. Thus, considering all the scenarios together and addressing them via a single platform is extremely difficult. Nonetheless, the proposed scheme can be extended to some scenarios that can accommodate multiple retailers as well. 

For example, consider a scenario in which multiple retailers with similar energy portfolios are available in the electricity market. In such a scenario, a prosumer cannot be engaged with multiple retailers at the same time and it needs to select one retailer for energy trading at any given time. In that case, at any time slot, each prosumer needs to select one retailer from the set of all available retailers first and then interact with it to trade energy in the P2P network following the proposed single retailer model. One example of how this can be done is shown in Fig. \ref{Fig:MultiRetailer}. According to this figure, prosumers who are interested in participating in P2P trading can submit their bids to the platform through which retailers obtain information about market price, total available supply and demand of energy within the community, and the bids submitted by the prosumers through the platform. Once all information is available to the retailers, each retailer makes some offers to the prosumers, such as its service charge and percentage of return of profit that each prosumer will receive for seeking their services for P2P trading and negotiates with the prosumers in an iterative fashion. Finally, once prosumers select their suitable retailers after negotiation, prosumers trade their energy with their selected retailers following the single retailer-based P2P model.

\begin{remark}
An underlying assumption of the proposed P2P framework is that the transfer of electricity within the network always satisfies the physical constraints imposed by the power system operator. This can be imposed, for example, by having all stakeholders interact with the distribution network service provider to ensure that their transacted energy does not violate the constraints \citep{Sousa_RSER_Apr_2019}. Other examples of studies that explain how P2P trading can be done without violating network constraints and can be adopted in the proposed scheme can be found in \citep{Chapman_TSG_2018} and \citep{Chapman_AE_2021}.
\end{remark}

Of course, both prosumers and the retailer could have improved their gains if they could independently participate in both markets. For example, if each prosumer could bid in the spot market, it does not need to rely on the retailer and thus the prosumer could save both the subscription fee and reap the entire revenue of selling its surplus at the spot market. However, today's market does not allow prosumers to participate to bid and sell electricity. Hence, they need to collaborate with the retailer for the additional benefit, which would make their total benefit much greater than just participating in the local P2P trading.

In the retailer-based market, on the other hand, the retailer makes substantially more revenue by selling the electricity to its customer at its chosen price. However, with the emergence of P2P trading, retailers are in the process of potentially discarded from that market model, as discussed by a large volume of P2P trading literature, e.g., see \citep{ Tushar_AE_Jan_2021}. The proposed model exhibit a case, in which retailer has an opportunity to participate in the P2P market and gain the trust of the participants by offering them services that would benefit prosumers more than existing local P2P trading schemes. Thus, retailers will not lose their current customer base and will be able to minimize the loss of revenue that could have taken place otherwise when P2P trading will be approved by regulators to be implemented in the electricity market at a large scale.

Thus, the additional profit streams make the proposed P2P strategy attractive to both the retailer and participating prosumers to implement for their sustainability in the future electricity market.

\section{Conclusion}The peer-to-peer framework proposed in this perspective provides a pathway by which retailers in today's electricity market can be integrated into decentralized local electricity trading as a core component of how the future electricity market will be designed and operated. This opens opportunities to make peer-to-peer energy sharing attractive to both prosumers and retailers without compromising their characteristics within peer-to-peer and retailer markets respectively. The proposed design represents a significant shift from the fundamental concept of the peer-to-peer market without retailer involvement, under which the role of a retailer is clearly defined and the benefit to the prosumers under such a new framework is articulated. By incorporating additional capabilities and services to its existing portfolio, retailers will be able to provide additional services to their customers in terms of bidding for their energy in the spot market and create supplementary revenue streams for households. Consequently, retailers will not lose their current customer base and will be able to minimize the loss of revenue that could have taken place otherwise when P2P trading will be approved by regulators to be implemented in the electricity market at a large scale. Although making such a change in the fundamental structure of peer-to-peer trading could be challenging, it will also be extremely valuable to convince regulators to allow large-scale implementation of such trading in the electricity market. 

To fully develop and implement the framework, interdisciplinary research across electricity engineering, power system economics, computer and social sciences, and psychology will be needed. Most importantly, different stakeholders of the electricity market such as prosumers, generators, peer-to-peer platform providers, market operators, retailers, and regulators need to closely work together to deliver such a comprehensive and effective system. As such, some promising directions for future research to make the proposed system a reality include:

\subsection{Information and communication technology}In contrast to the existing P2P trading scheme, novel methodology will be needed to integrate retailers within the P2P platform, such as blockchain-based platforms~\citep{Naveed_BlockChain_2019}, for participating in the market. Such integration will need to ensure that the participating retailer cannot access prosumers' private information about energy and storage, while at the same time, collaborate with the prosumers to enable them participate in the spot/retail market.

\subsection{Energy ethics} Enabling prosumers to participate in both the P2P and spot (or retail) market will clearly incentivize more prosumers to share their electricity in the market. However, it will open opportunities for prosumers to cheat about their electricity information in the P2P market for participating in the spot market for more benefit. Further, prosumers with greater production may have more freedom to increase their revenue by depriving relatively customers with lower surplus. Hence, the sharing mechanism should have strict regulation about energy ethics~\citep{EnergyEthics_2017} in place for sustainability of the proposed method.

\subsection{Artificial intelligence based approach for coordination} Forming a FPP and participating in the spot or retail market after local energy trading will be dictated by the prosumers' forecast about energy and price conditions. Hence, artificial intelligence-based sophisticated forecasting needs to be integrated into the P2P platform that will enable each prosumer of the network to take an informed and accurate decision.

\subsection{Comprehensive modelling of multiple retailer-based P2P trading}Once the importance and relevance of retailers' participation are established, more comprehensive frameworks to model scenarios with multiple retailers within the energy network will need to be developed. The model should be able to capture heterogeneous interest and interaction patterns of retailers within the market. For example, different retailers may have different types of energy contracts with renewable and non-renewable energy generators that can affect their business dealings with different customers with and without renewable energy resources. Thus, the model should be comprehensive and computationally efficient enough to consider all the scenarios together and addressing them via a single platform.

\subsection{Retailers' incentive design} While the proposed framework describes how retailers can be integrated into the P2P market, participating is such a market will clearly make retailers to loose revenue compared to existing retailer-based markets. Hence, suitable incentives need to be established for the retailers to ensure that participating in P2P market and the subsequent revenues are in the interest of the retailers. The benefit of participating in the spot/retail market by using prosumers' surplus should be rewarding for the retailers and hence the policy for such cases will require innovations.

\subsection{Region specific experimental validation} While the proposed single-retailer and multiple-retailer based frameworks exhibit the benefits of retailers' involvement in the P2P market, the deployment of the scheme would require approval by the energy regulators of regions where it is deployed. Since different regions can have different electricity market models, the proposed framework may need to be modified accordingly and experimentally validated before deployment in the energy market.

%\bibliography{Natureenergy}
%\bibliographystyle{apalike}

\end{document}